%% file: suzero_interspeech2020.tex
\documentclass[a4paper]{article}

\usepackage{INTERSPEECH2020}

\usepackage{graphicx}
\graphicspath{{fig/}}
\newcommand{\imagesep}{\vspace*{-4pt}}
\newcommand{\tablesep}{\vspace*{-6pt}}

\usepackage{booktabs}
\usepackage{tabularx}
\usepackage{multirow}
\newcommand{\mytable}{
    \centering
    \renewcommand{\arraystretch}{1.1}
    }

\newcolumntype{C}{>{\centering\arraybackslash}X}
\newcolumntype{L}{>{\raggedright\arraybackslash}X}
\newcolumntype{R}{>{\raggedleft\arraybackslash}X}
\newcolumntype{P}[1]{>{\raggedright\arraybackslash}p{#1}}
\usepackage{siunitx}
\usepackage{etoolbox}                           
\newcommand{\ubold}{\fontseries{b}\selectfont}  
\robustify\ubold                                

\usepackage{amsmath}
\usepackage{amssymb}
\DeclareMathOperator*{\argmin}{arg\,min}
\renewcommand{\vec}[1]{\boldsymbol{\mathbf{#1}}}

\usepackage{calc}
\usepackage{microtype}
\usepackage{url}
\usepackage{xcolor}

\newcommand\blfootnote[1]{\begingroup
                          \renewcommand\thefootnote{}\footnote{#1}
                          \addtocounter{footnote}{-1}
                          \endgroup}

\usepackage{cite}
\title{Vector-quantized neural networks for acoustic unit discovery \\ in the ZeroSpeech 2020 challenge}
\name{Benjamin van Niekerk \qquad Leanne Nortje \qquad Herman Kamper}
\address{E\&E Engineering, Stellenbosch University, South Africa}
\email{benjamin.l.van.niekerk@gmail.com, nortjeleanne@gmail.com, kamperh@sun.ac.za}

\usepackage[prependcaption,textsize=scriptsize]{todonotes}
\definecolor{mycolor}{HTML}{FF6600}

\begin{document}

\maketitle

\input{abstract}
\input{introduction}
\input{model_vqvae}
\input{model_vqcpc}

\input{experiments}
\input{conclusion}

\newpage
\bibliography{mybib}

\end{document}

%% file: abstract.tex
\begin{abstract}
In this paper, we explore vector quantization for acoustic unit discovery. Leveraging unlabelled data, we aim to learn discrete representations of speech that separate phonetic content from speaker-specific details. 
We propose two neural models to tackle this challenge -- both use vector quantization to map continuous features to a finite set of codes.
The first model is a type of vector-quantized variational autoencoder (VQ-VAE). 
The VQ-VAE encodes speech into a sequence of discrete units before reconstructing the audio waveform.
Our second model combines vector quantization with contrastive predictive coding (VQ-CPC). 
The idea is to learn a representation of speech by predicting future acoustic units.
We evaluate the models on English and Indonesian data for the \textit{ZeroSpeech 2020} challenge. 
In ABX phone discrimination tests, both models outperform all submissions to the 2019 and 2020 challenges, with a relative improvement of more than 30\%. 
The models also perform competitively on a downstream voice conversion task. 
Of the two, VQ-CPC performs slightly better in general and is simpler and faster to train. 
Finally, probing experiments show that vector quantization is an effective bottleneck, forcing the models to discard speaker information.
\end{abstract}

%% file: introduction.tex
\section{Introduction}

Modern speech and language technologies are developed with massive
amounts of annotated data. 
However, large datasets of transcribed speech are not available for {low-resource} languages and building new corpora can be prohibitively expensive.
As a result, tools like automatic speech recognition and text-to-speech are not available for many of the world's languages.

To address this problem, \textit{zero-resource speech processing} aims to develop methods that can learn directly from speech without explicit supervision. 
The goal is to leverage unlabelled data to discover representations that capture meaningful phonetic contrasts while being invariant to background noise and speaker-specific details.
These representations can then be used to bootstrap training in downstream speech systems and reduce requirements on labelled data. 
Additionally, since infants acquire language without explicit supervision, the discovered representations can be used in cognitive models of language learning~\cite{rasanen_speechcom12,schatz+feldman_ccn18,shain+elsner_naacl19}.

Over the last few years, progress in this area has been driven by the \textit{ZeroSpeech Challenges}~\cite{versteegh+etal_sltu16,dunbar+etal_asru17,dunbar+etal_interspeech19}.
\textit{ZeroSpeech 2020} consolidates previous challenges, allowing submissions to both the 2017 and 2019 tracks.
We focus on \textit{ZeroSpeech 2019: Text-to-Speech Without Text}, which requires participants to discover \textit{discrete} acoustic units from unlabelled data. 
From the discovered units, the task is then to synthesize speech in a target speaker's voice.
Synthesized utterances are evaluated in terms of intelligibility, speaker-similarity, and naturalness. 
While similar to voice conversion~\cite{kain+etal_icassp98,chou+etal_interspeech18}, an explicit goal of \textit{ZeroSpeech 2019} is to learn \textit{low-bitrate} representations that perform well on phone discrimination tests.
In contrast to work on continuous representation learning~\cite{zeghidour+etal_interspeech16,heck+etal_ieice18,chung+etal_interspeech19,wang+etal_icassp20,last+etal_spl20}, this encourages participants to find discrete units that correspond to distinct phones.\footnote{As a point of reference, phonetic transcriptions encode speech at a rate of about 39 bits per second~\cite{coupe2019different}.}

Early approaches to acoustic unit discovery typically combined clustering methods with hidden Markov models~\cite{varadarajan+etal_acl08,lee+glass_acl12,siu+etal_csl14,lee+etal_tacl15,ondel+etal_pcs16}.
More recent studies have explored neural networks with intermediate discretization~\cite{badino+etal_interspeech15,eloff+etal_interspeech19,chorowski+etal_taslp19,tjandra+etal_interspeech19}.
In this paper, we investigate vector quantized (VQ) neural networks for acoustic unit discovery, and propose two models for the \textit{ZeroSpeech~2020} challenge.

The first model is a type of vector-quantized variational autoencoder (VQ-VAE)~\cite{vandenoord+etal_neurips17}. 
The VQ-VAE  maps speech into a discrete latent space before reconstructing the original waveform.
Instead of using WaveNet~\cite{vandenoord+etal_arxiv16}, we opt for a lightweight recurrent network as the decoder.
The result is a smaller, faster model that can be trained on a single GPU.


Inspired by vq-wav2vec \cite{baevski+etal_iclr20}, the second model combines vector quantization with contrastive predictive coding (VQ-CPC).
Using a contrastive loss, the model is trained to distinguish future acoustic units from a set of negative examples.
We compare across-speaker and within-speaker sampling for negative examples and show that the latter is important for speaker invariance.


In ABX phone discrimination tests on English and Indonesian data, the models outperform all other submissions to the \textit{ZeroSpeech 2019} and \textit{2020} challenges.
On the voice conversion task, both models are competitive with VQ-CPC achieving the best naturalness and speaker-similarity scores on the English dataset.
Finally, in probing experiments, we analyze the effect of vector quantization, showing that it imposes an information bottleneck that separates phonetic and speaker content.

%

%% file: model_vqvae.tex
\section{Vector-quantized neural networks}


\subsection{Vector quantization}
\label{sec:vq-layer}

The VQ layer consists of a trainable codebook $\{e_1, e_2, \ldots, e_K \}$ with $K$ distinct codes. 
In the forward pass, a sequence of continuous feature vectors $z  := \left\langle z_1, z_2, \ldots, z_T \right\rangle$ is discretized by mapping each $z_i$ to it's nearest neighbor in the codebook. 
Concretely, we find $k := \argmin_j || z_i - e_j ||^2$ and replace $z_i$ with the code $e_k$, resulting in the quantized sequence $\hat{z} := \left\langle \hat{z}_1, \hat{z}_2, \ldots, \hat{z}_T \right\rangle$. 
Since the $\argmin$ operator is not differentiable, in the backward pass, gradients are approximated using the straight-through estimator \cite{bengio+etal_arxiv13}. 
To train the codebook, we use an exponential moving average of the continuous features. 
Finally, a \textit{commitment cost} is added to the loss to encourage each $z_i$ to commit to the selected code. 
For a more detailed explanation see \cite{vandenoord+etal_neurips17}.

\subsection{Vector-quantized variational autoencoder}
\label{sec:vqvae}

Our first model is a type of VQ-VAE inspired by the WaveNet autoencoders proposed in \cite{chorowski+etal_taslp19}. 
To reduce computational costs, we replace the WaveNet decoder~\cite{vandenoord+etal_arxiv16} with a lightweight RNN-based vocoder \cite{lorenzotrueb+etal_interspeech19}.
Together with automatic mixed precision~\cite{micikevicius+etal_iclr18}, this allows us to train on a single GPU.
Additionally, to learn a low-bitrate representation, we use a much smaller codebook.
Finally, we release code and pretrained weights.\footnote{\scriptsize \url{https://github.com/bshall/ZeroSpeech}}

\begin{figure}[!b]
    \centering
    \includegraphics[width=0.925\linewidth]{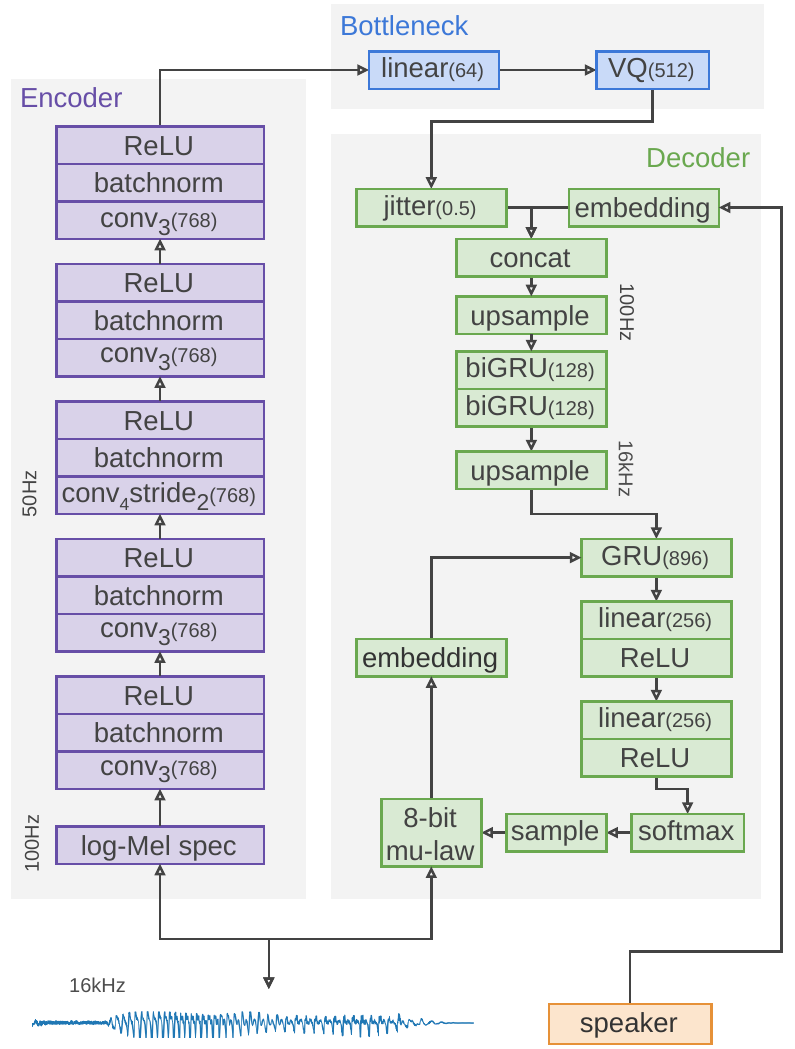}
        \imagesep
    \caption{
    VQ-VAE:
    A convolutional encoder (purple) takes a speech waveform as input and outputs downsampled continuous features.
    These are discretized (blue) using vector quantization.
    The decoder (green) then tries to reconstruct the input waveform from the discrete representation using an RNN-based vocoder conditioned on a speaker embedding.
    }
    \label{fig:vqwav}
\end{figure}

\label{sec:vqvae_model}

\textbf{Model description.}
The VQ-VAE can be divided into the three components shown in Figure~\ref{fig:vqwav}.
The \textit{encoder} takes a speech waveform sampled at 16 kHz as input and computes a log-Mel spectrogram.
The spectrogram is processed by a stack of 5 convolutional layers, which downsamples the input by a factor of 2.
In the \textit{bottleneck}, the output of the encoder is projected into a sequence of continuous features. 
 The representation is then discretized using a VQ layer with 512 codes.
Finally, the \textit{decoder} tries to reconstruct the original waveform. 
To predict the next sample, we condition an autoregressive model on the output of the bottleneck, the speaker identity, and past waveform samples.

For acoustic unit discovery, the VQ-VAE balances two opposing pressures. 
On the one hand, the encoder must preserve information from the input  to accurately reconstruct the waveform.
On the other hand, vector quantization imposes an information bottleneck, forcing a compressed representation that discards non-essential details.
To encourage the bottleneck to specifically discard speaker information, we condition the decoder on speaker identity during training.
\textbf{Training details.}
We train the model to maximize the log-likelihood of the waveform given the bottleneck, i.e.\ we minimize the sum of the reconstruction error and commitment cost:
\begin{equation*}
\mathcal{L} := - \frac{1}{N} \sum_{i=1}^N \log p(x_i| \hat{z}) + \beta { \frac{1}{T} \sum_{i=1}^T ||z_i - \text{sg}(\hat{z}_i)||^2},
\end{equation*}
where $\left\langle x_1, x_2, \ldots, x_N  \right\rangle$ is a sequence of waveform samples, $\beta$ is the commitment cost weight, and $\text{sg}(\cdot)$  denotes the stop-gradient operator.
The model is trained on minibatches of 52 segments, each 320~ms long.
We use the Adam optimizer \cite{kingma+ba_iclr15} with an initial learning rate of $4 \cdot 10^{-4}$, which is halved after 300k and 400k steps.
The network is trained for a total of 500k steps.

\textbf{Voice conversion.}
At test time, we can generate speech in a target voice by conditioning the decoder on a specific speaker.
First, we encode a source utterance into a sequence of acoustic units.
Since the bottleneck separates speaker details form phonetic information, we can replace the speaker while retaining the content of the utterance.
Specifically, the output of the bottleneck is concatenated with the target speaker embedding and piped to the decoder.

\textbf{Practical considerations.}
Since our goal is to discover phone-like representations, adjacent frames within the same phone should ideally be mapped to the same unit.
To encourage consistency across frames, we apply time-jitter regularization~\cite{chorowski+etal_taslp19}: during training the code assigned to each frame may be replaced by one of its neighbors, forcing the discovered codes to be useful across multiple time steps.
We apply jitter directly after the bottleneck, with a replacement probability of 0.5.
Another common issue with vector quantization is codebook collapse -- where only a few codes are ever selected~\cite{kaiser+etal_icml18, baevski+etal_iclr20}. 
We found that batch normalization, coupled with large batch sizes, improved codebook utilization.


%% file: model_vqcpc.tex
\subsection{Vector-quantized contrastive predictive coding}
\label{sec:vqcpc}

Contrastive predictive coding (CPC) is a recently proposed framework for unsupervised representation learning \cite{vandenoord+etal_arxiv18}.
Using a contrastive loss, models are trained to distinguish future observations from a set of negative examples.
The idea is that to make accurate predictions, the model must infer global structure in speech (e.g. phone identity) while discarding low-level details.

Recent studies have shown that CPC learns representations that capture phonetic contrasts and transfer well across languages \cite{riviere+etal_icassp20, kahn_icassp20}.
In this paper, we adapt CPC to the task of acoustic unit discovery by incorporate vector quantization.\footnote{\scriptsize \url{https://github.com/bshall/VectorQuantizedCPC}}
Previous work \cite{baevski+etal_iclr20, hadjeres+crestel_arxiv20} has also explored the combination of vector quantization with contrastive learning.
However, the goal in \cite{baevski+etal_iclr20} is to learn representations for a downstream automatic speech recognition.
We show that some design choices (e.g. product codebooks) that work well in this context are not ideal for acoustic unit discovery.



\begin{figure}[!t]
    \centering
    \includegraphics[width=0.975\linewidth]{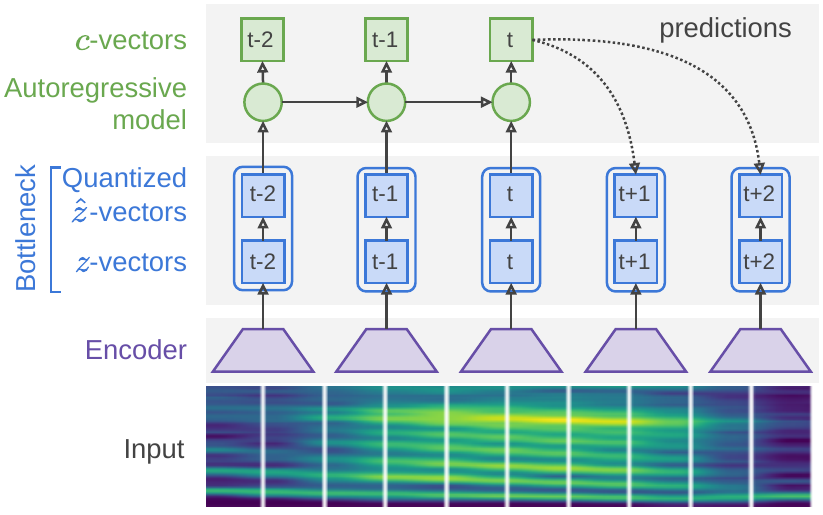}
    \imagesep
    \caption{
    VQ-CPC: An encoder (purple) encodes speech (parametrized as a log-Mel spectrogram) to a sequence of continuous vectors $z$. 
    Using a VQ bottleneck (blue) the $z$-vectors are quantized.
    The quantized $\hat{z}$-vectors are summarised by an autoregressive RNN (green) into context vectors $c$.
    Using this context, the model is trained to predict future codes.}
    \label{fig:vqcpc}
\end{figure}

\textbf{Model description.}
The VQ-CPC model is illustrated in Figure~\ref{fig:vqcpc}.
First, the \textit{encoder} maps input speech (parametrized as a log-Mel spectrogram) into a sequence of continuous features. 
The encoder consists of a strided convolutional layer (downsampling the input by a factor of 2), followed by a stack of 4 linear layers with ReLU activations.
Layer normalization is applied after each layer.
The \textit{bottleneck} is identical to the one described in \S\ref{sec:vqvae}. 
The output of the encoder is projected into a sequence of continuous latent vectors which are discretized using a VQ layer with 512 codes.
Finally, the \textit{autoregressive model} summarizes the discrete representations (up to time $t$) into a context vector $c_t$. 
Using this context, the model is trained to discriminate future codes from negative examples drawn from other utterances.

\textbf{Training details.}
Given a prediction horizon of $M$ steps, a trainable predictor matrix $W_m$, and a set $\mathcal{N}_{t,m}$  containing negative examples and the positive code $\hat{z}_{t+m}$, we minimize the InfoNCE loss~\cite{vandenoord+etal_arxiv18}:
\[
\mathcal{L}_t := - \frac{1}{M} \sum_{m=1}^M \log \left[ \frac{\exp(\hat{z}^\intercal_{t+m} W_m c_t)}{\sum_{\tilde{z} \in \mathcal{N}_{t,m}} \exp(\tilde{z}^\intercal W_m c_t)} \right].
\]
The 
loss is averaged over segments of 1.28 seconds and a VQ commitment cost is added.
We set the prediction horizon to $M = \textrm{6}$ steps and sample 17 negative examples per step.
We use the Adam optimizer, with a batches size of 64, and a learning rate of $4 \cdot 10^{-4}$. 
Each minibatch is divided into groups of 8 segments from which negative examples are sampled.
To address codebook collapse, we use a warm-up phase where we linearly increase the learning rate from $1 \cdot 10^{-5}$ over the first 150 epochs.

\textbf{Sampling negative examples.}
We investigate \textit{across-speaker} and \textit{within-speaker} sampling for negative examples.
In across-speaker sampling, negatives are drawn from a mix of speakers, while within-speaker sampling uses the same speaker.
We hypothesize that {within-speaker} sampling will encourage speaker-invariant representations since speaker information cannot be used to identify  the positive example.

\textbf{Voice conversion.}
VQ-CPC is not a generative model, so we train a separate vocoder on top of the discovered acoustic units for voice conversion.
The vocoder is similar to the decoder in Figure~\ref{fig:vqwav}, except the jitter layer is replaced with an embedding which reads in the code indices from the VQ-CPC bottleneck. Again, the target voice can be controlled by conditioning the vocoder on a specific speaker.

%% file: experiments.tex
\section{Experimental setup}
\label{sec:setup}

\begin{figure*}[t]
    \centering
    \includegraphics[width=0.99\textwidth]{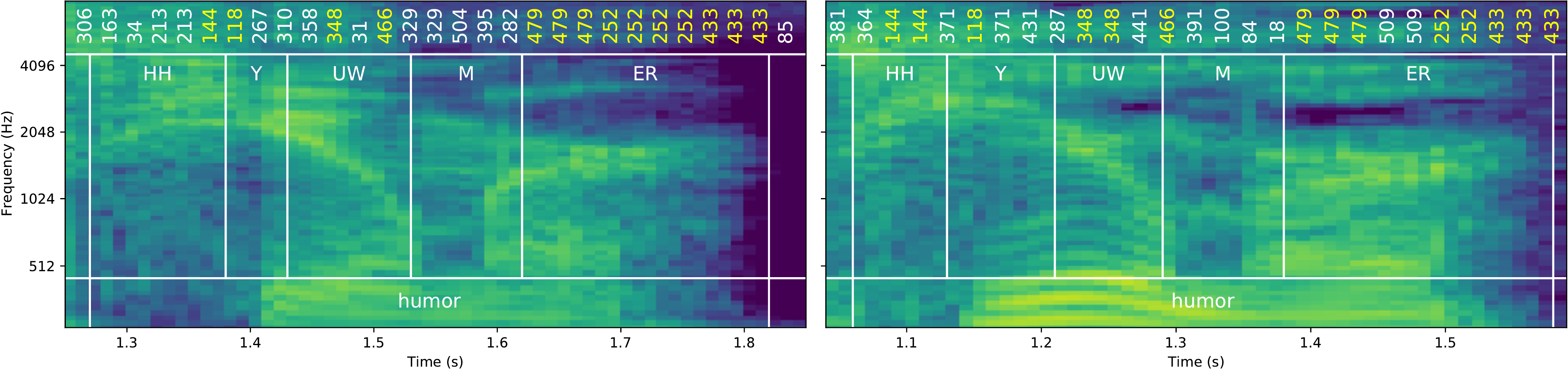}
        \imagesep
    \caption{
    The log-Mel spectrograms of speech segments taken from two different speakers. Overlaid are the aligned transcriptions and acoustic units from VQ-CPC. Common units in the two code sequences are highlighted in yellow.
    }
    \label{fig:codes}
\end{figure*}

\indent\textbf{Datasets.} 
We evaluate our models on the English and Indonesian datasets from the \textit{ZeroSpeech 2019 Challenge}.
Indonesian is a low-resource Austronesian language widely used as a lingua franca~\cite{sakti+etal_ococosda08,sakti+etal_tcast08}. 
Following the challenge guidelines, we use English as the development language. 
After tuning the models and hyperparameters on the English data, we apply the finalized training procedure to the Indonesian dataset.
For both languages, training data consists of about 15 hours of speech from over 100 speakers. 
An additional hour is provided per target speaker for voice conversion.
Finally, the test set contains approximately 30 minutes of speech from unseen speakers.

\textbf{ABX evaluation.} 
ABX phone discrimination tests are used to evaluate the discovered acoustic units~\cite{schatz+etal_interspeech13}.
The tests ask whether triphone $X$ is more similar to triphones $A$ or $B$.
Here, $A$ and $X$ are instances of the same triphone (e.g. ``beg''), while $B$ differs in the middle phone (e.g. ``bag'').
To measure speaker-invariance, $A$ and $B$ come from the same speaker, but $X$ is taken from a different speaker.
As a similarity metric, we use the average cosine distance along the dynamic time warping alignment path.
ABX is reported as an aggregated error rate over all pairs of triphones in the test set.

\textbf{Voice conversion.}
To assess voice conversion quality, human evaluators judge intelligibility, speaker-similarity, and naturalness.
For intelligibility, the evaluators orthographically transcribe the synthesized speech. 
A character error rate (CER) is then calculated by comparing the transcriptions to the ground truth.
Speaker-similarity and naturalness are scored on a scale from 1 to 5 (higher is better), with the latter reported as a mean opinion score (MOS).


\textbf{Baselines.}
The challenge baseline system combines a Dirichlet process Gaussian mixture model (DPGMM) for acoustic unit discovery~\cite{ondel+etal_pcs16} with a
parametric speech synthesizer based on Merlin~\cite{wu+etal_ssw16}.
The topline system feeds the output of a supervised speech recognition model to a text-to-speech system, both trained on ground-truth transcriptions.
See~\cite{dunbar+etal_interspeech19} for details.

We also include results for three other approaches. 
The first is the VQ-VAE-based system we submitted to the previous challenge~\cite{eloff+etal_interspeech19}, referred to here as VQ-VAE(spec).
Instead of generating audio waveforms directly, VQ-VAE(spec) uses a two-stage approach.
The model reconstructs log-Mel spectrograms, which are then fed to a separately trained FFTNet vocoder~\cite{jin+etal_icassp18} for synthesis.
Secondly, we include results for the system of Chen and Hain~\cite{chen+hain_interspeech20} -- one of the other top-performing submissions to \textit{ZeroSpeech~2020}.
Their system is similar to the WaveNet autoencoder of~\cite{chorowski+etal_taslp19}, but uses instance-norm layers in the encoder and adaptive instance normalization for speaker conditioning. 
In contrast to our models, Chen and Hain also downsample by a factor of 4 and use a large product codebook with $2^{16}$ codes.
Finally, we compare against vq-wav2vec \cite{baevski+etal_iclr20}.\footnote{We use the pretrained weights available at \scriptsize{ \url{https://github.com/pytorch/fairseq}}.} Note that vq-wav2vec was trained on the 960h Librispeech dataset.

\begin{table}[!t]
    \mytable
    \caption{Human and machine evaluations on the English and Indonesian test sets. 
        For MOS and similarity scores, higher is better. 
        For CER, ABX, and bitrate, lower is better.
        ABX scores for the discrete codes and auxiliary representations are shown under the ``code'' and ``aux'' columns respectively.
    }
    \tablesep
    \eightpt
    \begin{tabularx}{\linewidth}{@{}l@{\ \ }C@{\ \ }C@{\ \ }c@{\ \ }S[detect-weight,table-format=2.1]@{\ }S[detect-weight,table-format=2.1]@{\ \ }r@{}}
        \toprule
        & CER & MOS & Similarity & \multicolumn{2}{@{\ \ }c}{\underline{\ \ \ ABX (\%)\ \ \ }} & \\
        Model & (\%) & [1, 5] & [1, 5] & {code} & {aux} & {Bitrate} \\
        \midrule
        \underline{\textit{English:}} \\[3pt] 
        DPGMM-Merlin & 77 & 2.14 & 2.98 & 35.6 & {-} & \hphantom{0}72 \\      
        VQ-VAE(spec)~\cite{eloff+etal_interspeech19} & 67 & 2.18 & 2.51 & 27.6 & 23.0 & 173 \\
        Chen and Hain~\cite{chen+hain_interspeech20} & \ubold 18 & 3.61 & 2.57 & 20.2 & {-} & 386 \\
        vq-wav2vec~\cite{baevski+etal_iclr20} & {-} & {-} & {-} & 19.2 & 16.3 & 1205 \\[3pt]
        VQ-VAE & 39 & 3.62 & 3.49 & 14.0 & 13.2 & 412 \\
        VQ-CPC & 38 & \ubold 3.64 & \ubold 3.80 & \ubold 13.4 & \ubold 12.5 & 421 \\[3pt]
        Supervised & 43 & 2.52 & 3.10 & 29.9 & {-} & \ubold 38\\
        \midrule
        \underline{\textit{Indonesian:}} \\[3pt] 
        DPGMM-Merlin & 67 & 2.23 & 3.26 & 27.5 & {-} & \hphantom{0}75 \\
        VQ-VAE(spec)~\cite{eloff+etal_interspeech19} & 60 & 1.96 & 1.76 & 19.8 & 14.5 & 140 \\
        Chen and Hain~\cite{chen+hain_interspeech20} & \ubold 15 & \ubold 4.06 & 2.67 & 12.5 & {-} & 388 \\[3pt]
        VQ-VAE & 21 & 3.71 & 2.59 & \hphantom{~}6.2 & 8.3 & 424 \\
        VQ-CPC & 27 & 3.49 & 2.68 & \ubold 5.1 & \ubold 4.9 & 420 \\[3pt]
        Supervised & 33 & 3.49 & \ubold 3.77 & 16.1 & {-} & \ubold 35 \\
        \bottomrule
    \end{tabularx}
    \label{tbl:eval}
    \vspace{-6mm}
\end{table}

\section{Experimental results}
\label{sec:experiments}

Table 1 shows the evaluation results for the \textit{ZeroSpeech 2020 Challenge}.\footnote{The leader-board can be viewed at {\scriptsize \url{https://zerospeech.com/2020/results.html}}. Voice conversion samples for our models can be found at {\scriptsize \url{https://bshall.github.io/ZeroSpeech/}} and {\scriptsize \url{https://bshall.github.io/VectorQuantizedCPC/}}, respectively.}
On the ABX tests, our models outperform all other submissions to the 2019 and 2020 challenges.
We improve ABX scores on the English and Indonesian datasets by more than 30\% and 50\%, respectively.
On the English voice conversion task, VQ-CPC also achieves top naturalness and speaker-similarity scores, marginally beating the VQ-VAE. 
However on Indonesian, some of the other submissions perform better.
This discrepancy may be explained by a mismatch in the volume of our synthesized speech and the source utterances. On the English dataset, the volume difference is moderate at around 6.1 LUFS\footnote{Loudness Units relative to Full Scale (LUFS), see the \texttt{ITU-R BS.1770-4} standard.}. But on Indonesian, a larger disparity of 9.4 LUFS may have negatively impacted the results.

On intelligibility (CER), Chen and Hain~\cite{chen+hain_interspeech20} perform the best across both languages.
These results seem to indicate a trade-off between intelligibility and voice conversion quality. 
By using a larger codebook, Chen and Hain are able to improve CER at the cost of speaker-similarity.
A second trade-off, along a different axis, is bitrate against CER and ABX score.
While our models are able outperform the supervised topline, they operate at a much higher bitrate.

Comparing the two models, it is clear that the VQ-VAE and VQ-CPC perform similarly across all metrics.
However, VQ-CPC is an order of magnitude faster to train and was more robust to codebook collapse in our experiments.
A comparison to the VQ-VAE(spec) (from our previous submission~\cite{eloff+etal_interspeech19}), suggests that training an autoregressive decoder jointly with the encoder is beneficial.
Finally, it is interesting to note that our models (trained exclusively on unlabelled speech) achieve comparable ABX scores to the visually grounded VQ model of~\cite{harwath+etal_iclr20} -- which is trained on paired images and unlabelled spoken captions.


To show that the VQ bottleneck discards speaker information, we analyze the representations before and after quantization.
At each probe point, we train a multilayer perceptron with 2048 hidden units to predict the speaker identity.
We use mean-pooling after the non-linearity to aggregate features. 
Table~\ref{tbl:speaker_classification} shows the results of the probing experiments on English data.
Based on the drop in speaker classification accuracy across the probe points, the VQ layer clearly acts as an information bottleneck, forcing the models to discard speaker details.
Interestingly, CPC without vector quantization performs well on ABX tests but does not explicitly discard speaker information.
As a result, CPC alone was not capable of voice conversion in our experiments.
Table~\ref{tbl:speaker_classification} also compares within-speaker and across-speaker negative sampling for VQ-CPC (see \S\ref{sec:vqcpc}).
Within-speaker sampling results in better speaker invariance (lower speaker classification accuracy) and significantly lower ABX scores (13.4\% vs.\ 36.2\%).


To examine a few of the acoustic units discovered by VQ-CPC, Figure~\ref{fig:codes} plots two utterances along with the extracted codes.
The utterances are encoded to a similar sequence of units despite coming from different speakers.
Additionally, adjacent frames within a phone are often mapped to the same code.


\begin{table}[!t]
    \mytable
    \caption{
    Speaker classification results at probe points before and after quantization (shown under the ``pre-quant'' and ``code'' columns respectively). 
    }
    \tablesep
    \eightpt
    \begin{tabularx}{\linewidth}{@{}lCcCC@{}}
        \toprule
              & \multicolumn{2}{c}{Spkr. class. accuracy (\%)} & \multicolumn{2}{c}{ABX (\%)} \\
        \cmidrule(l){2-3}\cmidrule(l){4-5}
        Model & code & pre-quant & code & aux \\
        \midrule
        log-Mel spectrogram &  98.9 & - & 27.0 & - \\
        vq-wav2vec~\cite{baevski+etal_iclr20} & 76.4 & 98.7 & 19.2 & 16.3 \\
        VQ-VAE & 65.8 & 98.8 & 14.0 & 13.2 \\
        VQ-CPC (within) & 47.4 & 94.9 & 13.4 & 12.5 \\
        VQ-CPC (across) & 80.3 & 98.5 & 36.2 & 31.7\\
        CPC (within) & 99.7 & - & 16.4 & 13.8 \\
        \bottomrule
    \end{tabularx}
    \label{tbl:speaker_classification}
    \vspace{-6mm}
\end{table}

%% file: conclusion.tex
\section{Conclusions and future work}


We presented two neural models for acoustic unit discovery from unlabelled speech.
Using vector quantization, both models learn discrete representations of speech that capture phonetic content but discard speaker information.
They performed competitively on phone discrimination tests and a voice conversion task for the \textit{ZeroSpeech 2020} challenge.
Despite these merits, the models operate at high bitrates compared to phonetic transcriptions and a supervised topline.
In future work, we aim to lower bitrates and discover acoustic units that are more consistent across phones.\blfootnote{This work is supported in part by the National Research Foundation of South Africa (grant number: 120409) and a Google Faculty Award.}